\documentclass[aps,preprint]{revtex4}
\usepackage{epsfig}
\begin{document}
\def\I.#1{\it #1}
\def\B.#1{{\bf #1}}
\def\C.#1{{\cal  #1}}

\title{Quasi-Static Fractures in Disordered Media and Iterated Conformal Maps}
\author{Felipe Barra, H. George E. Hentschel$^*$, Anders Levermann and Itamar Procaccia}
\address{Dept. of Chemical Physics, The Weizmann Institute of Science, Rehovot, 76100, Israel\\
$^*$ Dept of Physics, Emory University, Atlanta Georgia}
\widetext
\begin{abstract}
We study the geometrical characteristic of quasi-static fractures in disordered media,
using iterated conformal maps to determine the evolution of the fracture pattern.
This method allows an efficient and accurate solution of the Lam\'e equations without 
resorting to lattice models. Typical fracture patterns exhibit increased ramification
due to the increase of the stress at the tips. We find
the roughness exponent of the experimentally relevant backbone of the fracture pattern; 
it crosses over from about 0.5 for small scales to 
about 0.75 for large scales, in excellent agreement with experiments. 
We propose that this cross-over reflects the increased ramification of the fracture pattern.
\end{abstract}
\maketitle
Considerable amount of theoretical work \cite{52Mus,86LL,99FM} on fracture in disordered media is based
on attempts to solve the equation of motion for an isotropic elastic body in the
continuum limit
\begin{equation} 
\rho \frac{\partial^2\B.u}{dt^2}=(\lambda+\mu)\B.\nabla(\B.\nabla\cdot \B.u)+\mu \nabla^2\B.u \ . \label{eqmot}
\end{equation}
Here $\B.u$ is the field describing the displacement of each mass point from its location
in an unstrained body and $\rho$ is the density. The constants $\mu$ and $\lambda$ are the Lam\'e constants.
In terms of the displacement field the elastic strain tensor is defined as
\begin{equation}
\epsilon_{ij}\equiv \frac{1}{2}\left(\frac{\partial u_i}{\partial x_j}+\frac{\partial u_j}{\partial x_i}\right) 
\ . \label{strain}
\end{equation}
For the development of a crack the important object is the stress tensor, which in linear elasticity is written as
\begin{equation}
\sigma_{ij}\equiv \lambda \delta_{ij}\sum_k\epsilon_{kk} +2 \mu \epsilon_{ij}
\ . \label{stress}
\end{equation}
When the stress component which is transverse to the interface of a crack exceeds a threshold value
$\sigma_c$, the crack can develop. When the external load is such that the transverse stress exceeds
only slightly the threshold value, the crack develops slowly, and one can neglect the second 
time-derivative in Eq. (\ref{eqmot}). This is the quasi-static limit, in which after each growth event
one needs to recalculate the strain field by solving the Lam\'e equation
\begin{equation}
(\lambda+\mu)\B.\nabla(\B.\nabla\cdot \B.u)+\mu \nabla^2\B.u =0 \ . \label{lame}
\end{equation}
In many previous works the problem was approached by discretizing 
Eq. (\ref{lame}) on a lattice \cite{90HR,81Sle,93ML,00PCP}. In this letter
we offer a novel approach based on iterated conformal maps; this method turned out to be very
useful in the context of fractal growth patterns \cite{98HL,99DHOPSS,00DFHP,00DLP} 
and it appears advantageous also for the present problem.

Although we can develop the approach in the full generality of Eq. (\ref{lame}), for the sake
of clarity in this Letter we will consider mode III fracturing for which a 3-dimensional elastic
medium is subjected to a finite shear stress $\sigma_{zy}\to \sigma_\infty$ as $y\to \pm \infty$.
Such an applied stress will create a displacement field $u_z(x,y)$, $u_x=0$,
$u_y=0$ in the medium. Despite the medium being three dimensional, therefore, the calculation of the strain and stress tensors
are two dimensional. 

We can describe a crack of arbitrary shape by its interface $\vec{x}(s)$,
where $s$ is the arc length which is used to parameterize the contour. 
We wish to develop a quasi-static model \cite{89BDL,90Ker} for the
 time development of this fracture in which discrete events advance the interface with a normal velocity 
\begin{equation}
\label{velocity}
v_n(s) = \alpha (\sigma_{zt}(s) - \sigma_c)
\end{equation}
if the transverse component of the stress tensor $\sigma_{zt}$ is greater than a critical yield value 
$\sigma_c$ for fracturing; otherwise no fracture propagation occurs. We will use the notation $(t,n)$ to describe
respectively the transverse and normal directions at any point on the two-dimensional crack interface.
Whenever the interface has more than one position $s$ for which $v_n(s)$ does not vanish, we will
respect the disorder by choosing the next growth position randomly with a probability proportional to $v_n(s)$.
This is similar to Diffusion Limited Aggregation (DLA) in which a particle is grown with a probability proportional
to the gradient to the field. One should note that another model could be derived in which all eligible fracture
sites are grown simultaneously, growing a whole layer whose local width is $v_n(s)$. This would be more akin to
Laplacian growth algorithms, which in general give rise to clusters in a different universality class than
DLA \cite{01BDP}. We prefer the first in order to take the disorder of the material explicitly
into account.

In mode III fracture $\B.\nabla \cdot \B.u=0$, and the Lam\'e equation
reduces to Laplace's equation
\begin{equation}
\label{laplace}
\partial^2 u_z/\partial x^2 + \partial^2 u_z/\partial y^2 = 0,
\end{equation}
and therefore $u_z$ is
the real part of an analytic function
\begin{equation}
\label{analytic}
\chi (z) = u_z(x,y) + i \xi_z(x,y)
\end{equation}
where $z = x+iy$. The boundary conditions far from the crack and on the crack interface can be used
to find this analytic function. It should be stated here that mode I and mode II fractures can
be reduced to a bi-Laplacian equation; then one needs to determine {\em two} rather than one
analytic functions. How to accomplish a growth model using iterated conformal maps for
those cases will be shown in a forthcoming publication \cite{01BLP}.

Far from the crack as $y \rightarrow \pm \infty$ we know $\sigma_{zy} \rightarrow \sigma_{\infty}$ or using the
stress/strain relationships Eq.~\ref{stress} we find that $u_z \approx [\sigma_{\infty}/\mu] y$.
Thus the analytic function must have the form \cite{2.3}
\begin{equation}
\label{far}
\chi (z) \rightarrow -i [\sigma_{\infty}/\mu] z \quad {\rm as}~ |z| \rightarrow \infty \ .
\end{equation} 

Now on the boundary of the crack the normal stress vanishes, i.e.
\begin{equation}
\label{near}
0=\sigma_{zn}(s) = \partial_n u_z=-\partial_t \xi_z .
\end{equation}
Since $\xi_z$ is constant on the boundary, we choose $\xi_z=0$, which in
turn is a boundary condition making the analytic function 
$\chi (z)$ real on the boundary of the crack:
\begin{equation}
\label{real}
\chi (z(s)) = \chi (z(s))^* \ .
\end{equation}

The direct determination of the strain tensor for an arbitrary shaped (and evolving) crack is
still difficult. We therefore proceed by turning to a mathematical complex plane $\omega$,
in which the crack is forever circular and of unit radius. The strain field for such a crack 
is well known, being the real part of the function $\chi^{(0)}(\omega)$ where
\begin{equation}
\label{solution}
\chi^{(0)} (\omega) = -i [\sigma_{\infty}/\mu](\omega - 1/\omega )
\end{equation}
This is the unique analytic function obeying the boundary conditions
$\chi^{(0)}(\omega) \rightarrow -i[\sigma_{\infty}/\mu]\omega$
as $|\omega| \rightarrow \infty$, while on the unit circle
$\chi^{(0)} (\exp i \theta ) = \chi^{(0)} (\exp i \theta)^*$.

Now invoke a conformal map $z = \Phi^{(n)}(\omega )$ that maps the exterior of unit circle in the mathematical plane
$\omega$ to the exterior of the crack in the physical plane $z$, after $n$ growth steps. This conformal map
is univalent by construction, and therefore admits a Laurent expansion
\begin{equation}
\Phi^{(n)}(\omega ) = F_1^{(n)}\omega + F_0^{(n)} +F_{-1}^{(n)}/\omega+F_{-2}^{(n)}/\omega^2+\cdots
\end{equation}
Then the
required analytic function $\chi^{(n)} (z)$ is given by the expression
\begin{equation}
\label{solution2}
\chi^{(n)}  (z) = -i [F_1^{(n)}\sigma_{\infty}/\mu]\Big({\Phi^{(n)}}^{-1}(z) - 1/{\Phi^{(n)}}^{-1}(z)\Big)
\end{equation}
From this we should compute now the transverse stress tensor:

\begin{eqnarray}
\label{transverse}
\sigma_{zt}(s) & = & \mu ~\partial_t u_z=\mu~ \Re \frac{\partial\chi^{(n)}(z)}{\partial s}\nonumber\\
&=&\mu ~\Re[\frac{\partial\chi^{(n)}(\Phi^{(n)}(e^{i\theta}))}{\partial \theta} \frac{\partial \theta}
{\partial s}]\nonumber\\&=&-\Re\frac{iF_1^{(n)}\sigma_{\infty} \frac{\partial}{\partial \theta}
(e^{i\theta}-e^{-i\theta})}{|\Phi^{'(n)}(e^{i\theta})|}\nonumber \\
&=&2 \sigma_{\infty}  F_1^{(n)}\frac{ \cos \theta}{|\Phi^{'(n)}(e^{i \theta} )|} \ ,
\end{eqnarray} 
 on the boundary. 
 
 Finally we describe how $\Phi^{(n)}(\omega)$ is obtained. Suppose that $\Phi^{(n-1)}(\omega)$
is known, with $\Phi^{(0)}(\omega)$ being the identity, $\Phi^{(0)}(\omega)=\omega$. We first compute
the transverse strain tensor  $\sigma_{zt}(\theta) = 
  2 \sigma_{\infty}F_1^{(n-1)}\cos \theta/|\Phi^{'(n-1)}(e^{-i \theta} )|$. In order
to grow according to the requirement (\ref{velocity}), we should choose growth sites
more often when $\Delta\sigma(\theta)\equiv \sigma_{zt}(\theta)-\sigma_c$ is larger. We therefore construct
a probability density $P(\theta)$ on the unit circle $e^{i\theta}$
which satisfies
\begin{equation}
P(\theta) = \frac{|\Phi^{'(n-1)}(e^{i \theta} )|\Delta\sigma(\theta)
\Theta(\Delta\sigma(\theta))}{\int_0^{2\pi}|\Phi^{'(n-1)}(e^{i \tilde\theta} )|\Delta\sigma(\tilde\theta)
\Theta(\Delta\sigma(\tilde\theta))d\tilde\theta}
\ , \label{weight}
\end{equation} 
where $\Theta(\Delta\sigma(\tilde\theta))$ is the Heaviside function, 
and $|\Phi^{'(n-1)}(e^{i \theta} )|$ is simply the Jacobian of the 
transformation from mathematical to physical plane. The next
growth position, $\theta_n$ in the mathematical plane, is chosen randomly with
respect to the probability $P(\theta) d\theta$. In this way the disorder of the
medium is mirrored in the growth pattern.
At the chosen position on the crack, i.e. $z= \Phi^{(n-1)}(e^{i\theta_n})$, we want to advance
the crack with a region whose area is the typical process zone for the material that we analyze.
According to \cite{90HR} the typical scale of the process zone is $K^2/\sigma^2_c$,
where $K$ is a characteristic fracture toughness parameter. Denoting the typical {\em area}
of the process zone by $\lambda_0$, we achieve growth with an auxiliary conformal map
$\phi_{\lambda_n,\theta_n}(\omega )$ that maps the unit circle to a unit circle with a bump
of area $\lambda_n$ centered at $e^{i\theta_n}$. 
An example of such a map is given by  \cite{98HL}:
\begin{eqnarray}
   &&\phi_{\lambda,0}(w) = w \left\{ \frac{(1+
   \lambda)}{2w}(1+w)\right. \nonumber\\
   &&\left.\times \left [ 1+w+w \left( 1+\frac{1}{w^2} -\frac{2}{w}
\frac{1-\lambda} {1+ \lambda} \right) ^{1/2} \right] -1 \right \} ^a \\
   &&\phi_{\lambda,\theta} (w) = e^{i \theta} \phi_{\lambda,0}(e^{-i
   \theta}
   w) \,,
   \label{eq-f}
\end{eqnarray}
Here the bump has an aspect ratio $a$, $0\le a\le1$. 
In our work below we use $a=2/3$.
 To ensure a fixed size step in the physical domain we choose
\begin{equation}
   \lambda_{n} = \frac{\lambda_0}{|{\Phi^{(n-1)}}' (e^{i \theta_n})|^2} \ .
   \label{lambdan}
\end{equation}
Finally the updated conformal map $\Phi^{(n)}$ is obtained as
\begin{equation}
\label{conformal}
\Phi^{(n)}(\omega ) = \Phi^{(n-1)}(\phi_{\lambda_n,\theta_n}(\omega )) \ . \label{iter}
\end{equation} 

The recursive dynamics can be represented as iterations
of the map $\phi_{\lambda_{n},\theta_{n}}(w)$,
\begin{equation}
   \Phi^{(n)}(w) =
\phi_{\lambda_1,\theta_{1}}\circ\phi_{\lambda_2,\theta_{2}}\circ\dots\circ
\phi_{\lambda_n,\theta_{n}}(\omega)\ . \label{comp}
\end{equation} 
Every given fracture is determined completely by the random itinerary
$\{\theta_i\}_{i=1}^n$. Eqs.(\ref{transverse}) together with (\ref{comp}) 
offer an analytic expression for the transverse stress field at any stage
of the crack propagation.

\begin{figure}
\centering
\includegraphics[width=.4\textwidth]{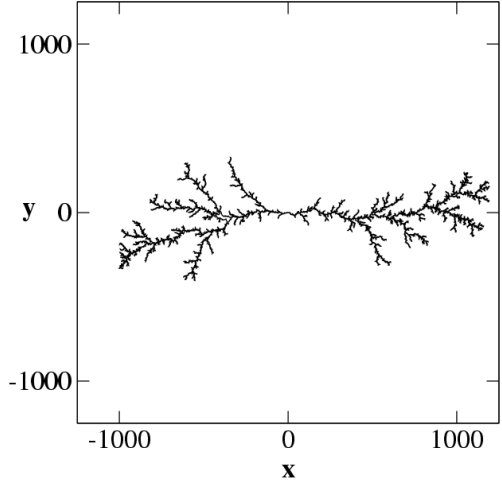}
\caption{A typical fracture pattern that is obtained from iterated conformal
maps. What is seen is the boundary of the fractured zone, which is the
mapping of the unit circle in the mathematical domain onto the physical domain.
Notice that the pattern becomes more and more ramified as the the
fracture pattern develops. This is due to the enhancement of the stress field
at the tips of the growing pattern}\label{fracture}
\end{figure}

Fig. \ref{fracture} exhibits a typical fracture pattern that is obtained with
this theory, with $\sigma_\infty=1$, after 10 000 growth events. The threshold value
of $\sigma_c$ for the occurrence of the first event (cf. Eq.(\ref{transverse}) is
$\sigma_c=2$. We always implement the first event. For the next growth event 
the threshold is $\sigma_c=2.9401...$.
We thus display in Fig.1 a cluster obtained with
$\sigma_c=2.94$, to be as close as possible to the quasi-static limit.
Nevertheless, one should observe that as the pattern develops, the
stress at the active zone increases, and we get progressively away from
the quasi-static limit. Indeed, as a result of this, for fixed boundary
conditions at infinity, there are more and more values of $\theta$ for which
Eq.(\ref{weight}) does not prohibit growth. Since
the tips of the patterns are mapped by ${\Phi^{(n)}}^{-1}$ to larger and larger
arcs on the unit circle, the support of the probability $P(\theta)$ increases,
and the fracture pattern becomes more and more ramified as the process
advances. The geometric characteristics of the fracture pattern are
{\em not} invariant to the growth. For this reason it makes little sense
to measure the fractal dimension of the pattern; this is not a stable
characteristic, and it will change with the growth. 

\begin{figure}
\centering
\includegraphics[width=.4\textwidth]{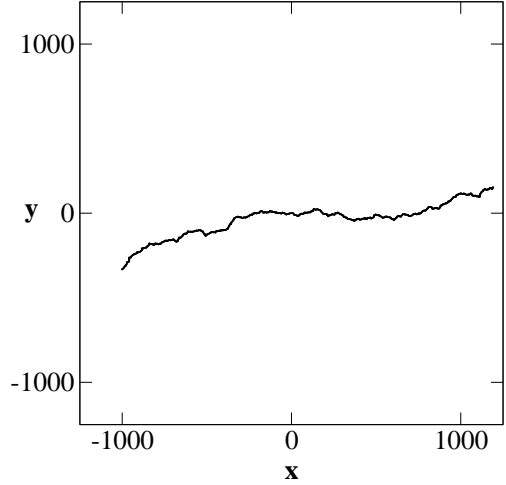}
\caption{A typical backbone of the fracture pattern. This is the 
projection onto the x-y plane of the experimentally
observed boundary between the two parts of the material that separate
when the fracture pattern hits the lateral boundaries}\label{backbone}
\end{figure}

On the other hand, we should realize that the fracture pattern is not
what is observed in typical experiments. When the fracture hits the boundaries
of the sample, and the sample breaks into two parts, all the side-branches of the 
pattern remain hidden in the damaged material, and only the backbone
of the fracture pattern appears as the surface of the broken
parts. The backbone does not suffer from the geometric variability discussed above.
In Fig. \ref{backbone} we show the backbone of the pattern displayed
in Fig. \ref{fracture}.

This backbone is representative of all the fracture patterns. we should
note that in our theory there are not lateral boundaries, and the
backbone shown does not suffer from finite size effects which may very
well exist in experimental realizations.

\begin{figure}
\centering
\includegraphics[width=.4\textwidth]{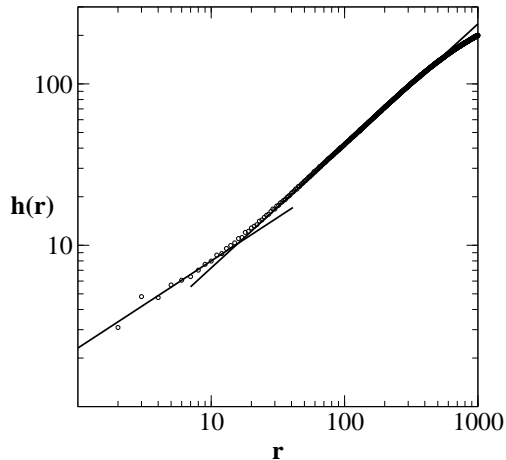}
\caption{$h(r)$ averaged
over all the backbone and over 70 fracture patterns each of which
of 10 000 fracture events. There is a cross-over between a scaling law
with roughness exponent $0.54\pm 0.05$ to and exponent of $0.75\pm 0.02$}
\label{scaling}
\end{figure}

In determining the roughness exponent of the backbone, 
we should note that a close examination of it reveals that {\em it is
not a graph}. There are overhangs in this backbone, and since we deal with
mode III fracturing, the two pieces of material {\em can} separate leaving
these overhangs intact. Accordingly, one should not approach the roughness
exponent using correlation function techniques; these may introduce serious
errors when overhangs exist \cite{95OPZ}. Rather, we should measure, for any given $r$,
the quantity \cite{97Bou}
\begin{equation}
h(r) \equiv \langle {\rm Max}\{y(r')\}_{x<r'<x+r}-{\rm Min}\{y(r'\}_{x<r'<x+r}\rangle_x \ .
\end{equation}
The roughness exponent $\zeta$ is then obtained from
\begin{equation}
h(r) \sim r^\zeta  \ ,
\end{equation}
if this relation holds. To get good statistics we average, in addition
to all $x$ for the same backbone,
over many fracture patterns. The result of the analysis
is shown in Fig. \ref{scaling}. 

We find that the roughness exponent for the backbone exhibits a clear
cross-over from 0.54 for shorter distances $r$ to 0.75 for larger
distances. Within the error bars these results are in excellent agreement 
with the numbers quoted experimentally, see for example \cite{97Bou}.
The short length scale exponent of order 0.5 is also in agreement with
recent simulational results of a lattice model \cite{00PCP} (which is by definition
a short length scale solution).
Bouchaud \cite{97Bou} proposed that the cross-over stems from transition between slow
and rapid fracture, from the ``vicinity of the depinning transition" to
the ``moving phase" in her terms. Obviously, in our theory we solve
the quasi-static equation all along, and there is no change of physics.
Nevertheless, as we observed before, the fracture pattern begins with
very low ramification when the stress field exceeds the threshold value
only at few positions on the fracture interface. Later it evolves to a much more ramified
pattern due to the increase of the stress fields at the tips of the
mature pattern. {\em The scaling properties of the backbone reflect
this cross-over}. We propose that this effect is responsible for
the cross-over in the roughening exponent of the backbone.
It is remarkable that in spite of the apparently
different mode of fracture (mode III is difficult to realize
experimentally) nevertheless the exponents appear extremely close
to those recorded experimentally for other fracture modes. This 
supports the conjecture
of universality proposed in \cite{97Bou}. 

We have thus demonstrated that iterated conformal maps offer an
efficient method for studying fracture patterns. Here we considered
only mode III quasi-static patterns. The theory for mode I and mode
II is available and will be presented elsewhere \cite{01BLP}. The generalization
to dynamical scaling, in which Eq.(\ref{eqmot}) is considered including
the time derivatives is akin to the transition from electrostatics
to electrodynamics. This is still an attractive goal for the road ahead.
\acknowledgments
We are indebted to S. Ciliberto for getting us interested in this
problem, and to J. Fineberg for some very useful discussions. This work has been supported in part by the
Petroleum Research Fund, The  European Commission under the
TMR program and the Naftali and Anna
Backenroth-Bronicki Fund for Research in Chaos and Complexity. A. L.
is supported by a fellowship of the Minerva Foundation, Munich, Germany.

\end{document}